\documentstyle[aps,pra,eqsecnum,preprint,tighten]{revtex}
\input{epsf}
\begin{document}

\def\lg{{\langle }}
\def\mt{\mapsto }
\def\od{\odot }
\def\ra{{\rightarrow }}
\def\Ra{{\Rightarrow }}
\def\rg{{\rangle }}
\def\srt{\sqrt{2}}
\def\vb{{\,|\,}}
\def\Tr{\hbox{Tr}}

\def\A{{\cal A}}
\def\B{{\cal B}}
\def\C{{\cal C}}
\def\D{{\cal D}}
\def\E{{\cal E}}
\def\F{{\cal F}}
\def\G{{\cal G}}
\def\H{{\cal H}}
\def\J{{\cal J}}
\def\P{{\cal P}}
\def\Q{{\cal Q}}
\def\W{{\cal W}}
\def\X{{\cal X}}
\def\Z{{\cal Z}}

\def\At{{\tilde A}}
\def\Bt{{\tilde B}}
\def\Ct{{\tilde C}}
\def\Dt{{\tilde D}}
\def\Ft{{\tilde F}}
\def\Mt{{\tilde M}}
\def\Pt{{\tilde P}}
\def\Qt{{\tilde Q}}
\def\Rt{{\tilde R}}
\def\St{{\tilde S}}
\def\Zt{{\tilde Z}}
\title{Choice of Consistent Family, and Quantum Incompatibility%
\thanks{Phys. Rev. A {\bf 57}, 1604 (March, 1998).}}

\author{Robert B. Griffiths\thanks{Electronic mail: rgrif@cmu.edu}\\
Department of Physics\\ Carnegie Mellon University\\ Pittsburgh, PA 15213,
U.S.A.}

%\date{Version of 8 October 1997}
\maketitle

%\begin{abstract}

	In consistent history quantum theory, a description of the time
development of a quantum system requires choosing a framework or consistent
family, and then calculating probabilities for the different histories which it
contains.  It is argued that the framework is chosen by the physicist
constructing a description of a quantum system on the basis of questions he
wishes to address, in a manner analogous to choosing a coarse graining of the
phase space in classical statistical mechanics. The choice of framework is not
determined by some law of nature, though it is limited by quantum
incompatibility, a concept which is discussed using a two-dimensional Hilbert
space (spin half particle).  Thus certain questions of physical interest can
only be addressed using frameworks in which they make (quantum mechanical)
sense.  The physicist's choice does not influence reality, nor does the
presence of choices render the theory subjective.  On the contrary, predictions
of the theory can, in principle, be verified by experimental measurements.
These considerations are used to address various criticisms and possible
misunderstandings of the consistent history approach, including its predictive
power, whether it requires a new logic, whether it can be interpreted
realistically, the nature of ``quasiclassicality'', and the possibility of
``contrary'' inferences.

%\end{abstract}

			\section{Introduction}
\label{intro}

	The consistent history approach 
\cite{gr84,gr87,gh90,om92,gr93b,om94,gr96} to
quantum theory provides a precise conceptual framework for describing how a
closed quantum system (isolated from its environment, or with the environment
itself included as part of a single closed system) develops in time.  It
reproduces the results of standard textbook quantum theory while avoiding the
well-known paradoxes associated with the ``measurement problem'', Bell's
inequality, and the like \cite{gr96,gr94}.  It shows promise for application to
quantum optics \cite{br97}, quantum computing \cite{gn96} and quantum
cryptography \cite{gn97}.

	At the same time, the consistent history approach has been the subject
of various criticisms 
\cite{de87,de89,de90,de95,dk95,dk96,kt96,kt97,kt97b} which center
upon the fact that the formalism allows the physicist to choose from a very
large number of alternative frameworks or consistent families of histories, all
of which are considered ``equally valid'' in the sense that no fundamental
physical law determines which family should be used in any given case.  Were
this ``freedom of choice'' the counterpart of gauge symmetry in classical
electromagnetism, it would be of no concern.  However, different frameworks are
often mutually incompatible in a manner which means that the use of one to
describe a given physical system precludes the use of another, that certain
questions can only be addressed if one uses the appropriate framework, and that
results from incompatible frameworks cannot be combined to form a single quantum
description.  The resulting conceptual problems have given rise to the claim
that consistent history quantum theory is incomplete and lacking in predictive
power \cite{dk95,dk96}, that it is logically incoherent and incompatible with
physical realism \cite{de87,de89,de90,de95}, or that the physicist's choice
(within this interpretation of quantum theory) must necessarily influence
reality \cite{de87,de95}.

	While all of these claims can be countered by a detailed analysis based
upon consistent history principles, as we shall see below, the fact that they
have been made by scientists who have studied consistent histories in some
detail means that the conceptual problems which give rise to them are likely to
trouble others as well.  Hence the present paper contains a detailed
examination of these problems, using a systematic formulation of the consistent
history approach published under the title ``Consistent histories and quantum
reasoning'', hereafter referred to as CHQR \cite{gr96}.  Indeed, the arguments
presented below are a continuation of a discussion already begun in CHQR as to
how the formal principles of consistent history quantum theory, which appear to
be sound, should be understood in physical terms and applied in particular
situations.  The basic conclusion of the present paper is that the criticisms
mentioned above do not indicate any deficiency or lack of logical coherence in
the consistent history approach, once it is properly understood.
Instead, the main problem is that the full import of the ``single consistent
family'' condition, which prohibits combining results from incompatible
families, has not always been appreciated, despite the fact that it was stated
in the very first publication devoted to consistent histories \cite{n1}, and
has been repeated many times since \cite{n2}.  In addition, even physicists who
accept the usual Hilbert space formulation of quantum mechanics often adopt,
without giving the matter enough thought, a mode of reasoning about its
physical consequences which is in basic conflict with the underlying
mathematics. A consequence is that one of the few interpretations of quantum
theory constructed in such a way that its rules of reasoning are consistent
with Hilbert space mathematics is criticized for being logically incoherent!

	The structure of this paper is as follows.  Section~\ref{example}
contains a brief review of the principles of consistent history reasoning as
found in CHQR, followed by applications of these principles to a simple
gedanken experiment.  Various consistent families, or frameworks, which are
incompatible with each other are constructed, and it is shown how they can be
used to derive a variety of physical conclusions.  The problems of choice and
of incompatibility emerging from this example are summarized at the end of the
section.

	The analysis of these two problems begins with a discussion of various
classical analogies, in Sec.~\ref{classical}.  Here it is argued that the
choice of a quantum framework is, among other things, like the choice of a
coarse graining for a classical phase space, and hence can properly be
considered a choice made by the physicist, not a consequence of some ``law of
nature''.  However, incompatibility between frameworks is a quantum effect with
no good classical analog, and must be dealt with in quantum terms.  This is the
topic of Sec.~\ref{incomp}, in which properties of a spin half particle are
used as a sort of quantum analogy in order to discuss general properties of
quantum incompatibility.  In addition, a simple example shows how the
predictions of consistent histories for what goes on in a closed quantum system
can, in principle, be confirmed by experimental measurements.

	Section~\ref{summ} begins with a summary of the conclusions which can
be drawn from the preceding arguments, and then continues with some
applications to various criticisms and (actual or potential) misunderstandings
of the consistent history approach.  These include the question of whether
consistent histories requires a new form of logic, whether it can be
interpreted in a realistic sense, a reply to the claim that the consistent
history approach lacks predictive power, comments on the nature of
``quasiclassicality'', and reasons why there cannot be a ``list of true
histories''.  The appendices contain somewhat more technical arguments which
address recent claims that consistent history quantum mechanics allows
``contrary inferences'' \cite{kt97,kt97b}, and that it cannot reproduce
consequences of standard quantum mechanics even if the future is thought to be
``quasiclassical'' \cite{kt96}.

		\section{A Simple Example of Quantum Reasoning}
\label{example}

	The basic structure needed in order to treat quantum mechanics as a
probabilistic theory is the same as for all other types of probabilistic
reasoning: a sample space of mutually exclusive elementary events, one and only
one of which occurs.  These ``events'' in the quantum case are {\it elementary
histories}; see Sec.~\ref{example} below for some examples.  They are analogous
to the classical histories obtained by flipping a coin several times in a row.
For example, the sample space for a coin flipped three times in a row consists
of eight sequences, HHH, HHT, \dots TTT, (H=heads, T=tails), one and only one
of which is actually realized when the experiment is carried out.

	Compound events, which are collections of elementary events, together
with the elementary events themselves, constitute the Boolean {\it event
algebra} to which probabilities are assigned.  (In the example of coin flipping
just mentioned, the four sequences corresponding to ``H on the first toss'' are
a compound event.)  In the quantum case, the algebra of events is called a {\it
consistent family} or {\it framework}.  It must be chosen according to
appropriate quantum mechanical rules, as discussed in CHQR.  Provided these
rules are satisfied, probabilities are assigned using a set of non-negative
weights calculated from the unitary time evolution engendered by
Schr\"odinger's equation.  Given these probabilities, the physical consequences
of the theory, typically in the form of conditional probabilities, are
calculated following precisely the same rules as in other applications of
standard probability theory.

	The peculiar features of quantum theory come about from the fact that
there are usually a large number of different ways of choosing the sample space
or consistent family, and it is necessary to pay careful attention to which of
these is being used to study a particular problem. In classical physics it
usually suffices to consider a single sample space, and in those instances in
which more than one is used, relating the different sample spaces is relatively
straightforward, unlike quantum theory, where the rules are more complicated.
For further details, see CHQR.

	A simple example which will serve to illustrate the principles of quantum
reasoning just discussed is shown in Fig.~1.  A photon which is initially in
channel $a$ passes through a beamsplitter $B$ and is later detected by one of
two detectors, $C$ and $D$.  The unitary time evolution produced by the
beamsplitter takes the form
\begin{equation}
  | a\rg\mt |s\rg = (|c\rg + |d\rg)/\srt,
\label{e2.1}
\end{equation}
where $| a\rg$ is a wave packet for the photon in the entrance channel $a$,
$|c\rg$ and $|d\rg$ are wave packets in the exit channels $c$ and $d$, and
$|s\rg$ is a coherent superposition of the latter. The interaction of the
photon with each of the detectors is given by the unitary transformations
\begin{equation}
  |c\rg|C\rg\mt |C^*\rg,\quad |d\rg|D\rg\mt |D^*\rg,
\label{e2.2}
\end{equation}
where $|C\rg$ indicates a detector in a state ready to detect a photon, and
$|C^*\rg$ the state which results when the photon is detected.  Note that
$|C\rg$ and $|C^*\rg$ differ from each other in some macroscopic, visible
property; for example, the position of a pointer.  The states
$|D\rg$ and $|D^*\rg$ of the other detector have a similar interpretation.
Some other particle, such as a neutron, could replace the photon without
altering the following discussion in any significant way.  Or one could imagine
a spin-half particle going through a Stern-Gerlach apparatus which separates an
$S_x=1/2$ initial state into two beams with $S_z=1/2$ and $S_z=-1/2$ directed
to separate detectors. 

	In the consistent history approach, the photon and the detectors are
thought of as forming a single, closed quantum system, and the quantum
mechanical description applies to the system as a whole. (In principle the beam
splitter should also be included, but since its quantum state plays no role in
the following discussion, it is omitted so as not to clutter up the notation.)
The unitary time development of the total system then takes the form
\begin{equation}
  |\Psi_0\rg \mt |s\,CD\rg \mt |S\rg,
\label{e2.3}
\end{equation}
where
\begin{equation}
  |\Psi_0\rg= |aCD\rg,
\label{e2.4}
\end{equation}
the initial state at a time $t_0$, evolves under the action of Schr\"odinger's
equation to a state $|s\,CD\rg$ at a time $t_1$ when the photon has passed
through the beam splitter, and then to
\begin{equation}
   |S\rg = (|C^*D\rg+|CD^*\rg)/\srt
\label{e2.5}
\end{equation}
at a time $t_2$ after the photon has been detected by one of the detectors.
Since $|C^*\rg$ and $|C\rg$ are macroscopically distinguishable, as are also
$|D\rg$ and $|D^*\rg$, $|S\rg$ is a {\it macroscopic quantum superposition}
(MQS) or ``Schr\"odinger's cat'' state.

	One can think of (\ref{e2.3}) as a {\it quantum history} based upon the
three times $t_0$, $t_1$, and $t_2$.  In the notation of CHQR it is represented
as a projector
\begin{equation}
  \Psi_0\od s\,CD\od S
\label{e2.6}
\end{equation}
on the tensor product space of histories, where we adopt the convention, here
and below, that a symbol outside a ket denotes the corresponding projector,
for example, 
\begin{equation}
  S= |S\rg\lg S|.
\label{e2.7}
\end{equation} 
For present purposes the tensor product structure is not important, and
(\ref{e2.6}) should simply be thought of as a statement that the system started
off in the initial state $\Psi_0$ (or $|\Psi_0\rg$) at $t_0$, was in a state 
$s\,CD$ at $t_1$, and in the state $S$ at time $t_2$.  The consistent history
approach supports such a ``realistic'' interpretation for reasons indicated in
CHQR and in Sec.~\ref{confirm} below. 

	The first consistent family or framework $\F_1$ we shall discuss has a
sample space of elementary families consisting of (\ref{e2.6}) together with
three additional histories
\begin{equation}
  	\Psi_0\od s\,CD\od \St,\quad
	\Psi_0\od (I-s\,CD)\od S,\quad
	\Psi_0\od (I-s\,CD)\od \St,
\label{e2.8}
\end{equation}
where $I$ is the identity operator on the Hilbert space, and
\begin{equation}
  \St = I- S,
\label{e2.9}
\end{equation} 
is the {\it negation} of $S$, and means ``$S$ did not occur''.  The four
projectors in (\ref{e2.6}) and (\ref{e2.8}) are the elementary histories which
constitute the sample space of $\F_1$. (They sum to $\Psi_0$ rather than to the
history space identity because we regard $\F_1$, along with the families we
shall consider later, as having a fixed initial event; see remarks towards the
end of Sec.~III in CHQR.)  However, quantum theory assigns a weight of zero to
the histories in (\ref{e2.8}), which means they are dynamically impossible, and
have zero probability to occur.  When discussing other families, we will
usually not display the zero-weight histories explicitly.

	Because the sample space has only one history, (\ref{e2.6}), with
finite weight, we can infer that
\begin{equation}
  \Pr(S_2\vb\Psi_0)=1,
\label{e2.10}
\end{equation}
where the subscripts, in this and later formulas, always refer to the times at
which the events occur: $S_2$ is a shorthand for ``$S$ at time $t_2$''.  Thus,
given that the system was in the state $|\Psi_0\rg$ at $t_0$, it will certainly
(with probability 1) be in the state $|S\rg$ at time $t_2$. As is well known,
unitary time evolution of this sort is something of an embarrassment for
standard quantum mechanics, because physicists typically prefer to draw the
conclusion that at time $t_2$ the system is in one of the two states $|C^*D\rg$
or $|CD^*\rg$, each with a certain probability, rather than in the
hard-to-interpret MQS state $|S\rg$.  The endless discussions about how to
interpret the state $|S\rg$ and resolve the corresponding ``measurement''
problem have been rightly criticized by Bell \cite{bl90}.

	The consistent history solution to this measurement problem is to
introduce a distinct consistent family $\F_2$ based upon a sample
space of elementary histories
\begin{equation}
  \Psi_0\od s\,CD\od C^*D, \quad \Psi_0\od s\,CD\od CD^*,
\label{e2.11}
\end{equation}
together with others of zero weight, such as $\Psi_0\od s\,CD\od C^*\Dt$.  The
physical interpretation of the first history in (\ref{e2.11}) is as follows: At
$t_0$ the photon was in state $|a\rg$ and the detectors were ready, at $t_1$
the photon was in $|s\rg$ and the detectors remained in the ready state,
whereas at $t_2$ detector $C$ had detected the photon, while $D$ was still in its
ready state, indicating that it had not detected the photon.  The second
history in (\ref{e2.11}) has the same interpretation with the roles
of $C$ and $D$ interchanged.

	A straightforward application of the rules for calculating
probabilities given in CHQR yields the results
\begin{equation}
  \Pr(C_2^*\vb \Psi_0) = 1/2 = \Pr(D_2^*\vb \Psi_0),
\label{e2.13}
\end{equation}
where, once again, the subscript $2$ indicates an event at the time $t_2$.  In
words, (\ref{e2.13}) tells us that, given the initial state $|\Psi_0\rg$ at
$t_0$, at time $t_2$ the probability is 1/2 that detector $C$ will have
detected the photon, and 1/2 that it will have been detected by $D$.
Similarly, one can show that
\begin{equation}
  \Pr(C_2^*\lor D_2^* \vb  \Psi_0) = 1,\quad 
	\Pr(C_2^*\land D_2^* \vb  \Psi_0) = 0,
\label{e2.14}
\end{equation}
where ``$\lor$'' stands for (non exclusive) ``or'', and ``$\land$'' means
``and''.  That is, at $t_2$ it is certain (probability one) that one or the
other of the detectors has detected the photon, and it is not true (because the
probability is 0) that both detectors have detected the photon. 

	Thus by using family $\F_2$, the consistent history approach provides a
definite, albeit probabilistic, answer to the question ``which detector
detected the photon?'', and this is the answer which everyone who uses standard
quantum theory agrees is correct.  Notice that this is done without the MQS
state $|S\rg$ ever entering the discussion and causing the sort of confusion
pointed out by Bell \cite{bl90}.  Indeed, according to consistent history rules,
$|S\rg$ at time $t_2$ {\it cannot} enter a discussion based upon $\F_2$,
because it is {\it incompatible} with the histories in $\F_2$ in the sense that
it cannot be added to $\F_2$ without violating the rules for
constructing quantum event algebras.  In particular, the event algebra must be
a Boolean algebra of projectors, and the projector $S$ does commute with either
$C^*D$ or $CD^*$, so cannot be part of a Boolean algebra which contains the
latter.  Combining any discussion of $|S\rg$ at time $t_2$ with $\F_2$ violates
the {\it single framework} rule of quantum reasoning, CHQR Sec.~III.  For
the very same reason, it is impossible to include either $C^*$ or $D^*$ at time
$t_2$ in family $\F_1$, and hence, if one employs $\F_1$, one cannot, according
to the rules of consistent history quantum theory, address the question of
which detector has detected the particle.  Such a discussion is ``meaningless''
in the technical sense that this interpretation of quantum theory assigns it no
meaning.  In the consistent history approach one cannot interpret $|S\rg$ to
mean ``$|C^*D\rg$ or $|CD^*\rg$ for all practical purposes'', or for any other
purposes.

	Quantum incompatibility is also the reason why we {\it cannot} combine
(\ref{e2.10}) with (\ref{e2.13}) to reach the conclusion
\begin{equation}
  \Pr(S_2\land C_2^*\vb \Psi_0) = 1/2,
\label{e2.15}
\end{equation}
a result which would follow immediately by standard probabilistic reasoning
from (\ref{e2.10}) and (\ref{e2.13}) {\it if these probabilities were based
upon the same sample space}. But the sample spaces for $\F_1$ and $\F_2$ are
distinct, and they cannot be combined to form a common sample space.  This is a
typical quantum mechanical effect which serves to define when two consistent
families are mutually incompatible.  Consequently, the left side of
(\ref{e2.15}) is meaningless, and assigning it the value 1/2, or any other
value, is a logical error.

	Now consider a third family of histories $\F_3$ based on a
sample space containing two histories of finite weight,
\begin{equation}
    \Psi_0\od c\od C^*, \quad \Psi_0\od d\od D^*,
\label{e2.16}
\end{equation} 
together with additional histories of zero weight.  (The projectors at $t_1$
and $t_2$ in the histories in (\ref{e2.16}), unlike those in (\ref{e2.11}), do
not project onto pure states. However, one could replace $c$ with $c\,CD$, $C^*$
with $C^*D$, etc. without altering the following discussion.) The family
$\F_3$ is obviously incompatible with $\F_2$ because the projectors $c$ and $d$
at $t_1$ do not commute with $s$, and it is obviously incompatible with $\F_1$
for the same reason, and also because of the detector states at $t_2$.

	The conditional probabilities (\ref{e2.13}) can be derived in $\F_3$ as
well as in $\F_2$; this reflects a very general result (CHQR Sec.~IV) that
conditional probabilities of this type always have the same values in any
framework in which they can be defined, whether or not the frameworks are
incompatible. In addition, in $\F_3$ (but not $\F_2$) one has
\begin{equation}
  \Pr(c_1\vb \Psi_0) = 1/2 = \Pr(d_1\vb \Psi_0),
\label{e2.17}
\end{equation}
a sort of microscopic analog of (\ref{e2.13}).  In words, given the initial
state at $t_0$, one can conclude that at time $t_1$ the photon will be either
in the $c$ channel with probability 1/2, or in the $d$ channel with the same
probability.  Note that these probabilities, unlike those of standard textbook
quantum theory, make no reference to measurements, which only take place some
time later.  That later measurements yield corresponding probabilities,
(\ref{e2.13}), is perfectly consistent with the general principle, see
Sec.~\ref{incomp}~D, that the probabilities assigned to events and histories by
the consistent history approach can always be checked, in principle, using
suitable (idealized) measurements.  (Measurements play no distinguished role in
consistent history quantum mechanics; if a closed quantum system contains some
measuring apparatus, the events involving the apparatus are analyzed in
precisely the same way as all other events.)

	Both (\ref{e2.13}) and (\ref{e2.17}) can be obtained by straightforward
application of the Born rule of standard quantum theory (although its use in
(\ref{e2.17}) might be regarded as problematical by some practitioners, since no
measurement is involved).  But the next result, 
\begin{equation}
    \Pr(c_1\vb \Psi_0\land C_2^*) = 1,\quad
	\Pr(d_1\vb \Psi_0\land C_2^*)=0,
\label{e2.18}
\end{equation}
goes beyond anything one can calculate with the Born rule.  It says that given
the initial state, and the fact that at $t_2$ it was detector $C$
which detected the photon, one can be certain that at time $t_1$ the photon was
in the $c$ channel and not in the $d$ channel.  Given the arrangement shown in
Fig.~1, this is a very reasonable result, and shows another way in which
consistent history methods make sense out of quantum measurements. That
appropriate measurements will reveal properties a measured system had before it
interacted with the apparatus is a principle used constantly in the design of
experimental equipment, and the inability of standard textbook quantum
mechanics to explain this is a defect just as serious as its inability to deal
with Schr\"odinger's cat.

	Notice that the result just discussed can be obtained using $\F_3$,
but not $\F_2$ or $\F_1$.  The question  ``was the photon in $c$ or
$d$ at time $t_1$?'' is not meaningful in $\F_1$ and $\F_2$ because
the projectors $c$ and $d$ needed to answer it are incompatible with these
families.  On the other hand, if we use  $\F_2$ we can derive the result
\begin{equation}
    \Pr(s_1\vb \Psi_0\land C_2^*) = 1.
\label{e2.19}
\end{equation}
That is, under the same conditions as in (\ref{e2.18}), one can conclude with
certainty that at time $t_1$, the particle was in the superposition state
$|s\rg$ defined in (\ref{e2.1}).  Just as in the case of
(\ref{e2.10}) and (\ref{e2.13}) discussed earlier, because (\ref{e2.18}) and
(\ref{e2.19}) were derived in two incompatible families, they cannot be
combined to reach the nonsensical conclusion
\begin{equation}
  \Pr(c_1\land s_1\vb \Psi_0\land C_2^*) = 1,
\label{e2.20}
\end{equation}
which, were it correct, would be the analog of asserting that a spin half
particle has both $S_x=1/2$ and $S_z=1/2$ at the same time.

	In summary, we have studied the gedanken experiment shown in Fig~1
using three separate and mutually incompatible frameworks or consistent
families, $\F_1$, $\F_2$, and
$\F_3$, each based upon its own sample space of histories involving events at
the three times $t_0$, $t_1$, and $t_2$, and have derived a number of different
(probabilistic) conclusions.  There are many (in fact, an uncountably infinite
number of) other frameworks which could have been employed to discuss the same
situation.  The fact that the same physical system can be discussed using many
different frameworks gives rise to the {\it problem of choice}: how does one
decide which framework is appropriate for describing what goes on in this closed
quantum system?

	A partial answer is suggested by the fact that certain questions of
physical interest, such as ``which detector detected the photon?'', and ``which
channel was the photon in before it was detected?'' can be addressed in certain
families but not in other families, because the projectors needed to provide a
quantum mechanical representation of the question are only compatible with
certain families. Thus incompatibility serves to limit choice, but at the same
time it gives rise to the {\it problem of incompatibility\/}: how are we to
understand the existence of, and the relationship among incompatible histories
and families?

	These two problems will be discussed in the following sections,
beginning with classical analogies in Sec.~\ref{classical}, which will help us
understand something of what is involved in choosing a quantum framework.
Unfortunately, there is no good classical analog of quantum incompatibility, so
the discussion of this concept in Sec.~\ref{incomp} is based upon simple
quantum examples.  Combining the results from these two sections leads to the
conclusions stated in Part A of Sec.~\ref{summ}.

			\section{Classical Analogies}
\label{classical}

	No classical analogy is fully adequate for understanding the
implications of quantum theory, which must, in the end, be considered on its
own terms.  Nevertheless, classical ideas can be extremely helpful in
suggesting fruitful ways to think about quantum principles.  This is
particularly true for the problem of choice, where classical analogies suggest
(though they obviously cannot prove) that the freedom a physicist has in
choosing a framework does not render the resulting theory unscientific or
lacking in objectivity, nor does it mean that the physicist's choice somehow
``influences reality''.

	As a first analogy, consider a professional historian composing a
history of Great Britain. Obviously, as a practical matter, he must choose to
emphasize certain subjects and ignore, or at least give much less
consideration, to others.  The choice of topics is determined by what interests
the historian (or perhaps the pressures on his professional career), and are
not the consequence of some ``law of history''.  Nor does his choice have any
influence on historical events (at least those occurring prior to when his work
is published).  Nevertheless, the account he produces, if he does his job well,
is not a purely subjective one, and other historians dealing with the same
subject matter ought to concur about the historical facts, even if they
disagree, as they surely will, about their significance.

	A second analogy is provided by a draftsman producing a representation
of a three-dimensional object on a two-dimensional piece of paper.  In order to
do this, he has to choose some perspective from which to view (or imagine
viewing) the object.  This choice is not determined by any ``law of
representation'', though it can be influenced by pragmatic considerations: some
things can be better represented using one perspective than another.  Nor,
obviously, does the choice influence the object itself, although it constrains
the amount and type of descriptive information which can be included in the
drawing.

	The choice of gauge in classical electromagnetism provides a third
analogy. This choice is a matter of convenience which obviously does not
influence the system being described.  In addition, the same physical
conclusions can be drawn using any gauge, so the choice is based entirely on
pragmatic considerations.  To be sure, this is different from what encounters
in the case of quantum families, since, as noted repeatedly for the example in
Sec.~\ref{example}, the topics one can discuss, and thus the conclusions one
can draw, do depend upon which family is employed.  Nonetheless, a particular
conditional probability can often be calculated in many different families, and
in all families in which it can be calculated it has the same value. Thus, for
example, (\ref{e2.13}) can be obtained using either $\F_2$ or $\F_3$, though
not $\F_1$.  Hence, relative to a particular conditional probability, the choice
among the families in which it can be computed is similar to the choice of a
gauge.
	
	Classical Hamiltonian mechanics, which describes a mechanical system
with $N$ degrees of freedom by means of a $2N$-dimensional phase space
$\Gamma$, provides a fourth analogy.  To be sure, once the phase space and the
Hamiltonian have been specified, the only remaining choice is that of the
appropriate phase space trajectory or orbit, $\gamma(t)$, and this is typically
made by assuming that the state $\gamma$ of the system is known at some time,
say $t=0$.  	While this choice bears some resemblance to the quantum choice
problem, the disanalogies outnumber the positive analogies.  Thus in the
quantum case, with the exception of histories involving only one time, there is
always more than one family consistent with a specification of the quantum
state at an initial time.  This reflects the fact that quantum mechanics is a
stochastic theory, unlike classical mechanics, which is deterministic.  In
addition, the choice between classical trajectories is a choice between
mutually exclusive alternatives: if one trajectory is a correct description of
the system, all others must be incorrect.  By contrast, the choice between
consistent families is never a choice between mutually exclusive alternatives
in this same sense, and while different quantum frameworks can be incompatible,
incompatibility is a very different relationship from that of being mutually
exclusive, as will be shown in Sec.~\ref{incomp}.  Nevertheless, it is
probably useful to include classical mechanics among our list of analogies, if
for no other reason than that the disanalogous features are sometimes
employed to analyze consistent history results, leading to the (erroneous)
conclusion that the latter are unsatisfactory.

	Classical {\it statistical\/} mechanics, itself a stochastic theory,
provides a better analogy for the quantum choice problem than does classical
mechanics.  Thus our fifth analogy is a coarse graining of the classical phase
space (CHQR, Sec.~VII~A), a covering of $\Gamma$ by a set of non-intersecting
cells, which provides a coarse-grained description of the system through
specifying which cell is occupied by the phase point at a succession of times.
(Making these a finite set of discrete times, and allowing the cells defining
the coarse graining to depend upon time, produces the closest analogy with the
quantum case.)

	Choosing a coarse graining and choosing a quantum framework are
analogous in that in both cases the choice is one made by the physicist in
terms of the physical problems he wishes to discuss, and is not something
dictated by the laws of nature.  At least in the case of the classical coarse
graining, it is at once obvious that this choice has no influence on the
behavior the system being described, although it may very well limit the type
of description which can be constructed.  For example, in order to describe
irreversible or other hydrodynamic behavior, one wants cells which are not too
large, but also not too small.  Because the coarse graining is chosen by the
physicist, the corresponding description can be said to be ``subjective'', but
this is a harmless subjectivity, because two physicists who use the same coarse
graining will reach identical conclusions.  There is another analogy in that
once a coarse graining has been specified, one and only one coarse-grained
history will actually occur, since a phase space trajectory will be in
precisely one cell of the coarse graining at any given time. In probabilistic
terms, a succession of cells from the classical coarse graining at a sequence
of different times is an element of a sample space, just as an elementary
history in a quantum framework is an element of a sample space, and given a
sample space, one expects that one and only one of the constituent ``events''
will actually occur.

	Along with the positive analogies just discussed, there is an important
disanalogy between descriptions based on classical coarse grainings and those
which employ quantum frameworks.  Two classical coarse grainings always possess
a {\it common refinement}, a third coarse graining constructed from the
intersections of the cells in the two collections.  Thus given any two
coarse-grained histories which describe the same system (over the same time
interval), there is always a finer-grained history constructed using the common
refinement, and which incorporates, or implies, each of the coarser histories.
The ultimate refinement of all coarse grainings is, of course, a description of
the system in terms of a phase space trajectory. Two {\it compatible} quantum
frameworks possess (as a matter of definition) a common refinement, and for
them the situation is quite analogous to the classical state of affairs just
discussed.  However, incompatible frameworks often arise in quantum theory, as
illustrated in Sec.~\ref{example}, and their mutual relationship is
necessarily different from that which exists among (always compatible)
classical coarse grainings.

	The solution to the problem of (quantum) choice suggested by the
classical analogies discussed above (with the exception of classical mechanics,
where a good analog of a quantum choice does not exist) is that the choice can
be made on pragmatic grounds, that is to say, relative to the physical problems
which concern the physicist who is constructing a quantum description.  There
is, to be sure, no guarantee that this solution is the correct one, for
analogies have their limitations, and we have just noted an important one in
the case of classical statistical mechanics (which also applies to the other
examples in this section): there is nothing which corresponds to quantum
incompatibility.  On the other hand, there is also no reason to suppose that,
just because they are imperfect, the analogies have led us to a false solution
of the choice problem. To obtain further insight, we need to take a closer look
at quantum incompatibility in quantum terms.  That is the subject of the next
section.

%\newpage

			\section{Quantum Incompatibility}
\label{incomp}
		\subsection{Logic in two dimensions}
\label{logic}

	In Sec.~\ref{example} we introduced three separate consistent families
of histories, or frameworks, to describe the same physical
system.  The analogy of the draftsman in Sec.~\ref{classical} suggests that
these can be thought of as three distinct, but equally valid, perspectives from
which to describe the system.  However, none of the analogies of
Sec.~\ref{classical} provides insight into the fact that $\F_1$, $\F_2$, and
$\F_3$ are incompatible alternatives. Quantum incompatibility of this sort has
no good classical analog, so one needs to employ quantum examples in order to
understand it.  The simplest non-trivial quantum system, a two-dimensional
Hilbert space, which we shall think of as representing the angular momentum
states of a spin half particle, can provide some useful insight into what is
going on. The discussion given here employs a somewhat different point of view
from that in CHQR Sec.~VI~A, and can be thought of as supplementing the latter.

	The histories of interest involve only one time, and the ``event'' at
this time corresponds to a {\it ray}, or one-dimensional subspace of the
Hilbert space, whose physical interpretation is that the component of spin
angular momentum in a particular direction in space is 1/2 (in units of
$\hbar$).  For example, a particular ray which we denote by $Z^+$ (the same
symbol denotes the ray and the projector onto the ray) corresponds to the
proposition ``$S_z=1/2$''.  What corresponds to the {\it negation} of this
proposition, to ``it is not the case that $S_z=1/2$''?  While one might suppose
it to be ``the spin is in some other direction'', standard quantum mechanics
identifies negation of a proposition with the orthogonal complement of the
corresponding subspace, which means that the negation of ``$S_z=1/2$'' is the
unique ray $Z^-$ perpendicular to $Z^+$, which corresponds to the proposition
``$S_z=-1/2$''.  Understanding negation in this way has a certain plausibility:
after all, if $S_z$ is not equal to 1/2, what other value can it have?
Furthermore, applying negation twice in a row brings us back to the original
property.  Nonetheless, the translation of logical terms, such as ``not'', into
a quantum context is a nontrivial matter, and the reader who is not content
with the traditional approach is welcome to try and improve upon it.

	If we agree that the negation of a ray $R$ in a two-dimensional Hilbert
space is the orthogonal ray $\Rt$, the next question to address is the
definition of ``$R$ {\it and} $S$'' ($R\land S$) for two rays.  Guided by the
rules of ordinary propositional logic, it is plausible that if $S=R$, then
$R\land R = R$, and if $S$ is the negation of $R$,
\begin{equation}
  \Rt\land R = R\land \Rt = 0,
\label{e4.3}
\end{equation}
where 0 is proposition which is always false, corresponding to a projector onto
the origin of the Hilbert space.  But what if $S$ is neither $R$ nor
$\Rt$?  We might suppose that in such cases there is always a third ray $T$
(which depends, of course, on $R$ and $S$) such that
\begin{equation}
  R\land S = T.
\label{e4.4}
\end{equation}
However, (\ref{e4.4}) is unsatisfactory, for consider the result
\begin{equation}
  \Rt\land T = \Rt\land(R\land S) = (\Rt\land R)\land S = 0\land S = 0
\label{e4.5}
\end{equation}
obtained by applying the usual logical rules governing ``$\land$''.  For this
to hold, $T$ must be the same as $R$, since in all other cases we are assuming
that $\Rt\land T$ is another ray, not 0.  However, the argument in (\ref{e4.5})
applied to $T\land \St$ tells us that this, too, is 0, so that $T$ must be the
same as $S$. But $T$ cannot be equal to both $R$ and $S$, since, by hypothesis,
they are unequal.

	One way out of this dilemma is to follow Birkhoff and von Neumann
\cite{bvn36} , and assume that whenever $R$ and $S$ are unequal, ``$R$ and
$S$'' is false:
\begin{equation}
  R\neq S \Ra R\land S= 0.
\label{e4.6}
\end{equation}
Taking the negation of both sides of the equality in (\ref{e4.6}), and using
the usual rules of logic, we conclude that, under the same conditions, ``$R$ or
$S$'' ($R\lor S$) is true:
\begin{equation}
  R\neq S \Ra R\lor S= I,
\label{e4.7}
\end{equation}
where $I$ represents the proposition which is always true, corresponding to the
identity operator on the Hilbert space.  Alas, (\ref{e4.6}) and (\ref{e4.7})
lead to a contradiction if we employ the standard distributive law of
propositional logic:
\begin{equation}
  Q\lor(R\land S) = (Q\lor R)\land (Q\lor S).
\label{e4.8}
\end{equation}
The reason is that if $Q$, $R$, and $S$ are three distinct rays, the left side
of (\ref{e4.8}), using (\ref{e4.6}) and assuming that $Q\lor0= Q$, is $Q$,
while the right side, see (\ref{e4.7}), is $I\land I=I$.  Birkhoff and von
Neumann's solution to this problem is to adopt a ``quantum logic'' in which
(\ref{e4.8}) no longer holds.

	Consistent histories has a very different way of escaping the dilemma
posed by (\ref{e4.5}).  When $S$ is different from either $R$ or $\Rt$, the
properties corresponding to these two rays are said to be {\it incompatible},
and $R\land S$ is considered to be {\it meaningless\/}: it is not a formula
composed according to the syntactical rules which govern meaningful quantum
discourse.  To illustrate the difference between this and the approach of
Birkhoff and von Neumann, suppose that $S_z=1/2$ is true.  According to
(\ref{e4.6}), ``$S_x=1/2$ and $S_z=1/2$'' is false, as is ``$S_x=-1/2$ and
$S_z=1/2$''.  The normal rules of logic would then tell us that both $S_x=1/2$
and $S_x=-1/2$ are false, which cannot be the case, since one is the negation
of the other.  Consequently, the usual rules of logical reasoning have to be
modified if one uses the Birkhoff and von Neumann system.  By contrast, in the
consistent history approach, ``$S_x=1/2$ and $S_z=1/2$'' is neither true nor
false; instead, it is meaningless.  Thus the truth of $S_z=1/2$ tells us
nothing at all about $S_x$, so we do not reach a contradiction.

	Similarly, in the consistent history approach it makes no sense to ask
``is $S_x=1/2$ or is $S_z=1/2$?'', for such a question implicitly assumes that
the two can be compared: one is true and the other false, or perhaps both are
true, or both are false.  But none of these is possible under the rules of
consistent history reasoning \cite{n3}.  Unless propositions belong to the same
framework, which in the particular example we are considering (but not in
general; see below and App.~A) is equivalent to requiring that the
corresponding projectors commute with each other, the consistent history
formalism allows no logical comparison between them, of any kind.  In
particular, it is very important to distinguish {\it incompatible}, which is
the relationship between $S_x=1/2$ and $S_z=1/2$, from {\it mutually
exclusive}, the relationship between $S_z=1/2$ and $S_z=-1/2$.  As already
noted, the truth of $S_z=1/2$ tells us nothing at all about the truth or
falsity of the incompatible $S_x=1/2$.  However, the truth of $S_z=1/2$ at once
implies that $S_z=-1/2$ is false, since the two are mutually exclusive: if a
spin half particle emerges in one channel from a Stern-Gerlach apparatus, it
surely does not emerge in the other channel.

	Standard textbook quantum mechanics escapes the dilemma of (\ref{e4.5})
by yet a different method, by ignoring it and treating a quantum system as a
``black box'' which can be subjected to external (classical?) measurements.
But ignoring a problem does not necessarily make it go away, and the enormous
conceptual confusion which besets ``the measurement problem'' in standard
quantum theory is an almost inevitable consequence of its inability to provide
clear principles for discussing the properties of a spin half particle.

%%\newpage

		\subsection{Generalization to histories}
\label{general}

	In CHQR, two frameworks or consistent families $\F$ and $\F'$ were
defined to be incompatible if they lack a common refinement, that is, if there
is no consistent family $\G$ which contains all the histories in both $\F$ and
$\F'$.  Consistent with that usage, we shall say that two histories $Y$ and
$Y'$ are incompatible if there is no (single) consistent family which contains
both of them, and a framework $\F$ and a history $Y'$ are incompatible if there
is no consistent family $\G$ which contains $Y'$ along with all the histories
in $\F$.  The spin half case discussed above is best thought of as a ``quantum
analogy'' which illustrates some, but not all of the features of
incompatibility.  In particular, it shows how incompatibility can arise when
the projectors representing properties at a single time do not commute with
each other. This is also the source of incompatibility for the histories and
families discussed in Sec.~\ref{example}~C, where the projectors on the Hilbert
space of histories (as in (\ref{e2.6}) and (\ref{e2.11})) do not commute with
each other. Incompatibility can also arise because, even though the relevant
projectors commute, consistency conditions are not satisfied for the enlarged
Boolean algebra.  For an example, see Sec.~VI~D in CHQR (and App.~A below).

	Whatever the source of incompatibility, consistent histories quantum
theory deals with it in the manner suggested above in the discussion of a spin
half particle, through a syntactical rule which states that all logical
argumentation must take place inside a single framework or consistent history
(for details, see CHQR).  On the other hand, the usual rules of reasoning apply
{\it within} a single framework, and for these purposes no modifications of
standard propositional logic are required.  And, just as in the case of a spin
half particle, it is important to distinguish {\it quantum incompatibility},
which means no comparison is possible between two histories (or results
calculated in two frameworks), from {\it mutually exclusive}, the relationship
between two elementary families belonging to the same framework.  Thus if $A$
and $B$ are mutually exclusive, the truth of one implies the falsity of the
other or, as histories, the occurrence of $A$ means that $B$ did not occur, and
vice versa.  If, on the other hand, $A$ and $B$ are incompatible, the truth of
one, which can only be defined relative to some framework which contains it,
tells us nothing about the truth or falsity of the other; they are ``not
comparable'': no comparisons between them are possible, because such
comparisons make no (quantum mechanical) sense.  In particular, if $A$ and $B$
are incompatible histories, the question ``did $A$ occur or did $B$ occur?''
does not make sense, for there is no framework in which they could be compared.

	We have already applied these principles in discussing the examples in
Sec.~\ref{example}, in particular when noting that no discussion of the MQS
state $S$ at time $t_2$ can be meaningfully combined with the histories
(\ref{e2.11}) used in $\F_2$ to discuss which detector detected the photon.
Similarly, the family $\F_1$ of unitary time evolution cannot be used to
discuss, much less answer, the question ``which detector?'', because $\F_1$ is
incompatible with histories containing $C^*$ or $D^*$ at $t_2$.  And since, as
noted above, incompatibility prevents us from asking whether the history $
\Psi_0\od s\,CD\od S$ occurred rather than $ \Psi_0\od s\,CD\od C^*D $ or $
\Psi_0\od s\,CD\od CD^*$, it does not make sense to ask which of the
incompatible frameworks $\F_1$ and $\F_2$ provides the correct description of
this gedanken experiment.

	Generalizing from this example, we conclude that whenever $\F$ and $\G$
are any two incompatible families, one cannot ask which provides the correct
quantum mechanical description; because the two sample spaces are mutually
incompatible, they cannot be compared.  But even if $\F$ and $\G$ are
compatible, it also makes no sense to ask which is correct, for in that case
their relationship is analogous to that of two classical coarse grainings, as
discussed in Sec.~\ref{classical}, and it is obvious that one cannot say that
one coarse graining provides a ``correct'' and another an ``incorrect''
physical description.  Since alternative frameworks are either compatible or
incompatible, we conclude that it is {\it never} sensible to ask ``which is the
correct framework?'', at least in the same sense in which one can ask ``which
is the correct classical phase space trajectory?''.  The relationship between
phase space trajectories is that between mutually exclusive possibilities,
whereas quantum frameworks are not related to each other in this way.  Of
course it does make sense to ask ``which framework is useful for solving this
particular problem?'', or ``which framework provides one with the greatest
physical insight?'', just as in the case of classical coarse grainings.

		\subsection{Using measurements to confirm histories}
\label{confirm}

	Further insight into how the descriptions provided by two or more
incompatible families are related to each other can be obtained by asking how
the predictions of consistent history quantum theory about the behavior of a
closed quantum system might, in principle, be verified experimentally.  Imagine
that the system is in a closed box, the initial state $\Psi_0$ is known, and
the consistent historian (who, needless to say, is outside the box \cite{n3b})
has carried out calculations of probabilities for different histories belonging
to a variety of distinct and mutually incompatible families.  How might these
probabilities be checked by measurements?  	Checking them involves opening
the box, or at least letting the contents interact in some way with the
external world, and this immediately raises the well-known problem that
measurements might disturb what is going on in the box, so that the results
they reveal might not be indicative of what would have happened in their
absence.  However, the very same objection can be raised in a ``classical''
context in which specifically quantum effects play no role.  A box might
contain sensitive photographic film which would be changed if one were to look
at it.  Or perhaps the box contains a bird which will escape if the box is
opened.

	In the end there is no ultimate proof that one's measurement is not
creating the effect observed, just as there can be no ultimate argument against
solipsism.  However, if he had a theory which was sufficiently powerful to
predict the perturbing effects of a measurement, and if other predictions of
this theory for other situations had been amply confirmed by experiment, a
physicist (in contrast to a philosopher) would probably be willing to use it in
order to design measurements which produced minimal perturbations, or to
calculate the effects of perturbations if these were unavoidable.  Consistent
history quantum mechanics is a theory of this sort, unlike standard textbook
quantum mechanics, whose inability to relate the results of measurements to
properties of the system being measured is one of its principal deficiencies.
Of course, using this approach cannot demonstrate that the consistent history
predictions of what is going on in a closed box are ``really true'', but it
does provide the sort of internally consistency which in other areas of science
(e.g., the study of the earth's core by means of seismic waves) is considered
good evidence for the existence of phenomena which cannot be directly observed.

	As an example, consider Fig.~2(a), which is a stripped down version of
the gedanken experiment in Fig.~1, as the detectors $C$ and $D$ have been
omitted from the quantum system enclosed in the box indicated by the heavy
line.  We consider two consistent families $\E_1$ and $\E_2$ with histories of
non-zero weight:
\begin{eqnarray}
  &&\E_1:\quad a\od c,\ a\od d
\label{e4.9}\\
  &&\E_2:\quad a\od s 
\label{e4.10}
\end{eqnarray}
referring to events at the times $t_0$ and $t_1$; the notation is that of
Sec.~\ref{example}. 

	To check the predictions of $\E_1$, we open holes in the sides of the
box, Fig.~2(b), at a time just before the photon would have encountered one of
the walls, to allow it to pass out of the box and be detected by one of the two
detectors $C$ and $D$.  Since the two histories in (\ref{e4.9}) are predicted
to occur with probability 1/2, checking this requires repeating the experiment
a number of times.  To determine whether opening the holes and placing
detectors outside the box has somehow perturbed the system, we employ a
consistent history analysis of the larger closed system which includes the
original box and the detectors $C$ and $D$ (along with the mechanism for
opening the holes, etc.).  For this larger system we ask: if the photon is
detected by $C$, was it earlier (while the holes were still closed) in the $c$
channel? The answer is affirmative; see (\ref{e2.18}) and the associated
discussion.  As in each realization of the experiment the particle is detected
by $C$ or $D$, this confirms the usual consistent history interpretation of the
two histories in (\ref{e4.9}) as mutually exclusive alternatives, one or the
other of which must occur in any particular case.

	If we employ family $\E_2$, the single history in (\ref{e4.10}) is
predicted to occur every time the experiment is carried out.  To verify this
requires the slightly more complicated arrangement shown in Fig~2(c). Once
again, the holes are opened at the last possible instant, but then a pair of
mirrors and a second beam splitter are used to compare the phases of the wave
packets emerging in the $c$ and $d$ channels.  Assume the path lengths are
arranged so that a photon initially in $a$ eventually emerges in $f$.  The fact
that the photon is always detected by $F$, and never by $E$, can then be used
as an experimental confirmation that it was in the state $s$ at $t_1$, since a
consistent history analysis of the larger closed system yields the result
\begin{equation}
  \Pr(s_1\vb \Psi_0\land F_2) =1.
\label{e}
\end{equation}
Once again, it is the ability of consistent histories to verify that a measured
system had a particular property before the measurement took place which lends
plausibility to the experimental confirmation of a prediction referring to
processes going on in a closed system.

	Checking events at intermediate, rather than final times in a
family of histories can be carried out using analogous, though somewhat more
complicated arrangements.  One can imagine the box to be supplied with ports
through which probes of appropriate type can be inserted at appropriate times.
The argument that this sort of measurement is possible in principle (that is,
without violating the laws of quantum theory), and that appropriate
measurements do not perturb the system in unacceptable ways, will be found in
Sec.~5 of \cite{gr84}.  Alternatively, one can suppose that a single closed box
contains both the system of interest and appropriate measuring apparatus which
interacts with the system at suitable times and records the results.  At the
end of the experiment the box is opened and the records are read.  In either
case, the consistent history approach allows one to discuss whether the system
actually did have the properties which the measurements revealed, and whether
these measurements perturbed the system in unacceptable ways.

	Whether the events of interest occur at intermediate or final times,
checking the predictions given by different incompatible frameworks always
involves alternative experimental arrangements which are either mutually
exclusive, as in Fig.~2(b) and (c), or of a sort in which one set of
measurements makes it impossible to discuss the system using some alternative
consistent family.  In any case, just as there is no single ``correct'' choice
of consistent family for describing the system, there is no single arrangement
of apparatus which can be used to verify the predictions obtained using
different families.

	Note that neither the framework nor the experimental arrangement needed
to check its predictions is singled out by some ``law of nature''.  Indeed,
were there some such law which told us, for example, that $\E_1$ is the correct
framework, we would have great difficulty interpreting the results of
measurements of the sort shown in Fig.~2(c).  In this sense, the physicist's
freedom to choose a framework, as suggested by the classical analogies in
Sec.~\ref{classical}, is not only possible in the presence of quantum
incompatibility, but appears necessary in order to have a theory with a
consistent physical interpretation.

			\section{Summary and Some Applications}
\label{summ}
		\subsection{Overall summary}
\label{over}
	
	The example in Sec.~\ref{example} together with the classical analogies
in Sec.~\ref{classical} and the discussion of incompatibility in
Sec.~\ref{incomp} lead to the following conclusions relative to the problems of
choice and incompatibility as stated at the end of Sec.~II.  There are many
posable consistent families or frameworks which can be used to describe the
same quantum system, and the choice of which of these to employ is a choice
made by the physicist based upon pragmatic considerations, namely the type of
physical problem he is trying to study or the question he wants to answer.
Because of quantum incompatibility, a question such as ``where was the photon
before it was detected?'' can be answered in certain frameworks and not in
others.  Given a particular physical question there are still, in general, a
large number of frameworks in which it can be discussed, but since they all
yield the same values for the relevant conditional probabilities, choosing
among them is, relative to this particular question, rather like the choice of
gauge in classical electromagnetism: the answer does not depend upon the
framework.

	The choice among alternative {\it compatible} families is closely
analogous to the choice among coarse grainings of a classical phase space, a
situation in which it is transparently obvious that the physicist's choice of a
mode of description has no influence whatsoever on the physical system being
described.  Furthermore, such a choice is not ``subjective'' in any
unacceptable way, for two physicists who use the same coarse graining will
reach identical conclusions. That these same conclusions are still correct in
the case of a choice between {\it incompatible} consistent families is not so
obvious, because there is no good classical analogy.  Nonetheless, their truth
is supported by the discussion of quantum incompatibility in Sec.~\ref{incomp}.
To begin with, the relationship of incompatibility between frameworks (or
between histories, etc.) is quite distinct from that between mutually exclusive
alternatives (as in a sample space), where the correctness of one means that
the others are necessarily false.  Incompatible frameworks are never related in
this way; and from the fact that some history has occurred, one cannot conclude
that some other history incompatible with it has not occurred.  Understanding
the difference between mutually exclusive and incompatible helps prevent one
from supposing that the physicist's choice of a framework somehow influences
the world by ``preventing'' or ``interfering with'' histories which occur in
some other framework.  Instead, the choice simply limits the type of
description which the physicist can construct.

	Additional support is provided by the discussion, in Part C of
Sec.~\ref{incomp}, of how predictions of what goes on in a closed system based
upon the consistent histories formalism, can in principle be checked
experimentally.  This ``operational'' point of view confirms that the choice of
consistent family is up to the physicist, but that the experimental
arrangements needed to confirm the predictions made using a particular
framework depend upon that framework.  By contrast, assuming that the framework
is determined by some ``law of nature'', rather than chosen by the physicist,
is rather unhelpful, since it produces new conceptual difficulties.

	Incompatibility, as noted in Sec.~\ref{incomp}~A, is a specifically
quantum concept which represents one way of meeting the (necessary!) process of
modifying the logic of classical propositions in order to achieve descriptions
which conform to the way standard quantum theory employs Hilbert space.  In the
consistent history approach, incompatibility is a syntactical rule which
governs which combinations of propositions about a quantum system can be said
to be physically meaningful.  In this respect it is quite different from the
approach of Birkhoff and von Neumann \cite{bvn36}, and has the advantage that
{\it within} a single consistent family there is no need to replace the
ordinary logic of propositions with a new ``quantum logic''. (By contrast,
textbook quantum mechanics in effect evades the logical issues by constructing
a phenomenological theory of measurement, thus giving rise to an insoluble
``measurement problem'' completely absent from the consistent histories
approach.)

	To be sure, quantum incompatibility and the choice among incompatible
frameworks for describing a particular quantum system are not matters which can
be easily understood using an intuition trained largely by the ``classical''
world of everyday experience.  This should come as no surprise: quantum
mechanics is distinctly different from classical mechanics, and it is only to
be expected that it contains concepts which conflict with pre-quantum thinking.
The same is true of special relativity in relationship to Newtonian mechanics.
The fact that events which occur at the same time in one coordinate system need
not be simultaneous in another, and the physicist's ability to change the time
difference between them by adopting a new coordinate system, have not been
considered insuperable barriers, or even serious objections to adopting special
relativity as a good scientific theory of the world, and supposing that it
provides a better description of physical reality than pre-relativistic
physics.  Although the analogy is, of course, not perfect, there seems to be no
reason why the physicist's liberty to choose a consistent family renders
consistent histories an unsatisfactory interpretation of quantum theory, nor
why quantum incompatibility should not be included with our other ideas about
what constitutes physical reality.

%\newpage

		\subsection{Logic, truth, and reality}
\label{truth}

	The structure of quantum reasoning set forth in CHQR, which is largely
compatible with earlier work by Omn\`es \cite{om92,om94}, agrees with but also
differs from ordinary propositional logic, depending upon one's point of view.
As long as the discussion is confined to a single framework (or ``logic'', in
Omn\`es' terminology) the usual rules employed for ordinary probabilistic
theories apply, and no new logical concepts are required.  In this respect one
can agree with Omn\`es that consistent histories quantum theory does not
require a new ``unconventional logic'' \cite{n4}.

	On the other hand, the syntactical rules which determine what
propositions can be part of a framework, whether two frameworks are compatible,
and the like, are decidedly non-classical, since their very definitions are
based upon properties of the quantum Hilbert space and the corresponding
unitary time transformations.  As these rules are essential for correct quantum
reasoning, and are thus central to the logical structure of quantum theory, it
is also correct to say that, as least in this respect, consistent histories
does involve a new form of logic, one with no classical counterpart. Thus
d'Espagnat's assertion \cite{n5} that consistent histories reasoning ``\dots
could only be valid in some as yet unspecified logic, of which it is not even
known how it could be self-consistent'' is not off the mark, though it is now
out of date, since in CHQR the required self-consistent logic has been
specified in considerable detail, filling in various items lacking in the
earlier \cite{gr93b}. As noted in Sec.~\ref{logic}, the use of some form of
non-classical reasoning is virtually inescapable once one accepts the association
of propositions, and their negations, with subspaces of Hilbert space, in the
way generally employed in standard quantum theory.  Whereas consistent
histories handles this in a very different way from the quantum logic of
Birkhoff and von Neumann, it still employs non-classical (and hence
counterintuitive) ideas.

	One important difference between Omn\`es and CHQR is in the definition
of ``true''.  In CHQR, ``true'' is interpreted as ``probability one''.  Thus if
certain data are assumed to be true, and the probability, conditioned upon
these data, of a certain proposition is one, then this proposition is true.
The advantage of this approach is that as long as one sticks to a single
framework, ``true'' functions in essentially the same way as in ordinary logic
and probability theory.  However, because probabilities can only be discussed
within some framework, comparisons of ``true'' between incompatible frameworks
are impossible, and in this sense ``true'' interpreted as ``probability one''
must be understood as relative to a framework.

	The feature just mentioned has been criticized by d'Espagnat
\cite{de87,de89,de95}.  But it is hard to see how to get around
it if one wishes to maintain (as do Omn\`es and I) that reasoning inside a
single framework should follow classical rules, and classical rules associate
``true'' (in a probabilistic theory) with ``probability one''.
Omn\`es' attempt to develop an alternative definition of ``true''
\cite{om92,om94} did not succeed \cite{dk96} , as he himself admits
\cite{ompc}, and at present there seems to be no serious alternative to CHQR.
In defense of the latter it is worth noting that, as a very general
principle, one cannot expect to import classical concepts into quantum theory
``duty free'', that is, without alterations, either in formal definitions, or
in intuitive properties, or both.  This is widely accepted by physicists in the
case of dynamical quantities: we do not insist that quantum position and
momentum commute with each other.  That it also holds for logical properties
and relationships deserves to be more widely appreciated; see the discussion of
``not'' and ``and'' in Sec.~\ref{logic} (and of ``contrary'' in App.~A).  Thus
it is unreasonable to expect that ``quantum truth'' will coincide with
``classical truth'' in every respect.  The definition adopted in CHQR has some
virtues; these include the fact that it is (almost) the same as its classical
counterpart in the case of a single framework, and reduces to the usual sense
of ``true'' in the classical (correspondence) limit in which incompatibility
disappears and all frameworks possess a common refinement.  Given these
properties, it seems reasonable to adopt the definition given in CHQR as a
suitable quantum counterpart of ``true'', at least until some superior
alternative appears on the scene.

	Similarly, d'Espagnat \cite{n6} thinks that consistent history quantum
theory cannot be interpreted in a realistic way, because it does not conform to
``a basic requirement of traditional realism, that there are facts that are
true quite independently of the conventions we decide to make as to which
consistent family of histories we prefer to discuss.'' In response, it is worth
noting that some modifications of traditional realism are only to be expected
if quantum theory is one of the truly revolutionary developments of twentieth
century science.  Nevertheless, the changes in pre-quantum realism required if
we adopt consistent histories are perhaps less radical than d'Espagnat's words
might suggest.

	Note, first of all, that the choice of framework by a physicist is not
something by which he can, in any sense, render true propositions untrue, or
make untrue propositions true.  The reason is that as long as frameworks are
compatible, choosing one or another is as ``harmless'' as selecting one of the
alternative coarse grainings of classical phase space, as discussed in
Sec.~\ref{classical}.  If, on the other hand, some framework is chosen which is
incompatible with a proposition, this choice does not make the proposition true
or false; instead the proposition is ``indiscussible'' within this framework.

	As this point is often misunderstood, it may be helpful consider the
specific example discussed in Part~C of Sec.~\ref{incomp}.  Suppose that
theorist T, standing outside the box of Fig.~2, has used family $\E_2$,
(\ref{e4.10}), to make a prediction, and experimentalist E has set up the
apparatus to test it using the arrangement in Fig.~2(c).  Suppose 
a second theorist, T*, carries out calculations using the incompatible family
$\E_1$ instead.  Will this alter what is going on inside the box?  Obviously
not; E's confirmation of T's prediction will be entirely unaffected by
T*'s calculation!  On the other hand, T* will, if he limits himself to
$\E_1$, be unable to provide a coherent account of how E's measurement confirms
T's prediction.  In order to do that, T* needs to employ $\E_2$.  If he does
so, he will, of course, obtain the same result as T.  On the other hand, if T*
can persuade E to carry out an alternative experiment using the arrangement of
Fig.~2(b)---this cannot, of course, be done at the same time as the experiment
in Fig.~2(c), but might be done later using a second photon---then this, or at
least a sufficient number of experiments of this type, will confirm T*'s
result, a situation which T can understand if he, too, employs $\E_1$.

	As well as showing that the choice of framework does not imply some
mysterious influence of mind upon matter, this example illustrates some other
aspects of quantum reality as viewed using consistent histories.  The existence
of a choice of framework does not render quantum theory subjective, for if T
and T* adopt the same framework, they come to the same conclusion.  Predictions
of what is going on in a quantum system can, in principle, be checked
experimentally; the theorist's freedom to choose a framework merely means that
alternative experimental arrangements are needed to check predictions made
using different frameworks.

	Objective descriptions (in the sense just discussed), experimental
confirmation, and the absence of peculiar influences of mind over matter are
aspects of traditional realism which continue to hold true in the quantum realm
if we accept the results of consistent histories.  But there is also something
very different: one cannot simultaneously employ $\E_2$ and $\E_1$ for
describing what is going on in the closed box.  Not because using one of them
makes the other false, but because there is no way of combining the results
from the two incompatible families.  Thus a ``unicity'' always present in
classical physics, the ability to combine any two descriptions of the same
system in a single description, is absent in quantum theory.  If one wishes to
maintain that this unicity is truly indispensable to realism, then, obviously,
consistent history quantum theory cannot be said to be ``realistic''.  The
alternative, which of course I favor, is to include quantum incompatibility as
part of our understanding of what quantum reality is all about, and why it
differs from what reality was thought to be like before the advent of quantum
theory.

		\subsection{Is consistent histories a predictive theory?}
\label{predict}

	As the result of a detailed study \cite{dk95,dk96}, Dowker and Kent
conclude that the consistent history approach to quantum theory lacks
predictive power, and for this reason is not satisfactory as a fundamental
scientific theory.  They accept the idea that once a consistent family
(``consistent set'' in their terminology) is specified, one and only one of the
corresponding elementary histories will take place, and quantum theory can only
assign a probability to the different possibilities.  That the theory is
probabilistic in this sense is not what concerns them.  Rather, it is the fact
that the consistent history approach, as a fundamental theory, treats all
frameworks or consistent families ``democratically'', and provides no criterion
to select out one in particular.  In other words, there is no ``law of nature''
which specifies the framework.  Thus, from their perspective, there is no way
of calculating probabilities of specific histories, since probabilities cannot
be computed without using some framework, and the theory does not tell one
which framework to use.

	Both CHQR and the present paper agree with Dowker and Kent that
consistent histories, as a fundamental theory of nature, does not single out a
particular framework.  The difference is that Dowker and Kent regard this
situation as unsatisfactory, whereas from the perspective presented here there
does not seem to be any problem if one regards the physicist's choice of
framework to be like the draftsman's choice of a perspective for representing a
three-dimensional object, or like the choice of a coarse graining of classical
phase space.  That is, the choice is dictated by the problems which the
physicist seeks to address, the questions he is attempting to answer, and not
by some law of nature.  Given some question of physical interest, quantum
incompatibility severely limits the choice, as noted in the case of the
specific example in Sec.~\ref{example}, and discussed further in
Sec.~\ref{incomp}.  There seems to be no reason why an element of choice of
this type should render a theory unsatisfactory.  It may not agree with certain
aesthetic criteria for what constitutes ``good science'', but such criteria
tend to be somewhat subjective, as in the case of Einstein's preference for a
deterministic rather than a probabilistic quantum theory.

	Insofar as Dowker and Kent regard the choice of framework as being a
choice among mutually exclusive alternatives, which seems to be their
perspective in Sec.~5.6 of \cite{dk96}, their point of view is quite different
from that presented above in Sec.~\ref{incomp}: that if two frameworks are
compatible, the relationship between them is analogous to that between
classical coarse grainings, and if they are incompatible, in the quantum sense
of that term, it is still not correct to suppose that the use of one framework
excludes the other in the same way that employing one trajectory in a classical
phase space excludes all other trajectories.  Thus one should never think of
one framework as ``correct'' and another as ``incorrect'', other than in a
sense like ``this is the correct framework to address the problem which I have
in mind''.

	A particular instance of the lack of predictive power, according to
Dowker and Kent, is that consistent history quantum theory cannot answer the
question, ``Will the world be quasiclassical tomorrow?''  See CHQR, Sec.~VII~B,
for a discussion of quasiclassicality in relationship to the ideas considered
here.  The term ``quasiclassical'' is somewhat vague, but the essence of Dowker
and Kent's concern lies in the fact that the consistent histories formalism
does not rule out the possibility of MQS states.  In particular there are
families in which only ``normal'' (non-MQS) states occur up to some particular
time, and MQS states appear at later times, so that ``quasiclassical'' behavior
up to some time is no guarantee that it will continue.

	The example in Sec.~\ref{example} can serve to illustrate this point.
Family $\F_1$ contains the ``non-quasiclassical'' MQS state $S$ at $t_2$,
whereas $\F_2$ has ``normal'' detector states $C^*D$ and $CD^*$; at the earlier
times $t_0$ and $t_1$ these families are identical, and the coherent
superposition state $s$ of the photon at time $t_1$, because it is easily
achieved and detected in the laboratory, can be considered compatible with a
``quasiclassical'' description.  The consistent history approach gives no
reason to choose $\F_2$ rather than $\F_1$ to represent the state of affairs at
$t_2$.  On the other hand, as pointed out in Sec.~\ref{incomp}, a question of
the form ``will the system be in $S$, or will it be in one of the two states
$C^*D$ or $CD^*$ at $t_2$?'' is not meaningful, because it requires, at least
implicitly, a comparison between mutually incompatible alternatives, and
quantum incompatibility implies that no such comparison is possible, that is,
it does not make sense.  It is like asking, ``does the spin half particle have
an $x$ or a $z$ component of angular momentum?''

	Note that ``quasiclassical'', interpreted as ``non-MQS'', is not a
quantum mechanical property as such, since it is not associated with a subspace
of the quantum Hilbert space.  Thus while both $|C^*D\rg$ and $|CD^*\rg$ in the
example of Sec.~\ref{example} refer to non-MQS states, the smallest subspace
which contains them also contains the MQS state $|S\rg$.  This suggests that
``quasiclassical'' is best thought of as a term belonging to the metalanguage
of quantum descriptions, the language used to discuss these descriptions,
rather than as a term which can itself enter into a quantum description.  To
use a classical analogy, the term ``large cells'' could be employed to
characterize a coarse graining of a classical phase space, and it belongs to
the metalanguage, for it obviously is not correct to think of ``large
cells'' as a property of the physical system itself.  The question ``will the
world be quasiclassical tomorrow?'' is thus comparable to ``will tomorrow's
coarse graining use large cells?''.  Both might make sense as part of a
discussion among physicists as to how to construct a description of a physical
system which best addresses the problems which interest them; neither refers
directly to properties of the system being described.

	To summarize, while it makes sense to compute probabilities of
different histories {\it given} a quasiclassical framework, it does not make
sense to assign a probability to such a framework, or treat it as one of a
collection of mutually exclusive physical alternatives.  Quasiclassicality is a
property of quantum descriptions, not a property of quantum systems, and the
question ``is this quantum system quasiclassical?'' is not meaningful, at least
when understood in the same sense as ``is the energy between 9 and 10 ergs?''.
For the same reason, quasiclassicality cannot be a condition in a quantum
probability, which is why a recent argument by Kent, examined in App.~B, is
inconsistent with the rules of quantum reasoning given in CHQR.

	Despite the obvious differences between the present paper and the
approach of Dowker and Kent, there is a sense in which the conclusions
complement each other.  By studying the structure of consistent histories under
the assumption that alternative consistent families are somewhat analogous to
the mutually exclusive possibilities represented by a sample space, Dowker and
Kent concluded that this approach does not result in a satisfactory scientific
theory.  That is perfectly compatible with the perspective of the present
paper.

	A rather different approach to consistent histories was considered by
Dowker and Kent in Sec. 5.4 of \cite{dk96} under the title ``many histories''.
See the following section for some comments.

		\subsection{List of histories}
\label{list}

	Given any consistent family of histories for a particular system, one
and only one of the elementary histories will actually occur. This statement
makes it tempting to suppose that it is possible to construct a list $\{
\F_j,F_j\}$ which assigns to every consistent family $\F_j$ a history
$F_j$, understood as the history which actually occurred in a particular system
or a particular realization of an experiment.  This temptation should be
resisted; no such list exists, at least if it is interpreted in the manner just
suggested.

	Note that if the $\F_j$ are alternative coarse grainings of a classical
phase space, the existence of such a list is not in doubt, for one simply takes
whatever classical trajectory actually occurs, and employs it to generate the
history $F_j$ by noting which cell of $\F_j$ is occupied by the trajectory at
each time.  In the same way, if the quantum list contains only consistent
families which are mutually compatible with one another, the list can be
constructed by using the common refinement.

	When, however, incompatible families occur in the list---as will
necessarily be the case if all consistent families are included---the quantum
list does not make sense.  Consider the example in Sec.~\ref{example}.  For
$\F_1$ there is only one history, (\ref{e2.6}), with positive weight, which
must therefore be $F_1$.  For $\F_2$ there are two histories with positive
weights, (\ref{e2.11}), so $F_2$ must be one of these.  But whichever it is, it
makes no sense to say that a single system (and a single experimental run) can
be correctly described by both $F_1$ and $F_2$, since at $t_2$ the detectors
cannot be said to be in the MQS state $S$ and also in one of the states $C^*D$
or $CD^*$.

	To be sure, one might adopt an alternative interpretation of the list
$\{ \F_j,F_j\}$, and understand it as referring to what happens in distinct but
nominally identical systems, or in successive repetitions of an experiment,
labeled by the subscript $j$.  In such a case there is no problem, for it is
not necessary to use the same framework every time an experiment is carried
out, or for describing different systems.  But this is obviously a very
different interpretation of the list from that employed above.

	Dowker and Kent's ``many histories'' interpretation, Sec.~5.4 of
\cite{dk96}, involves such a list, although their interpretation of what it 
means is not very clear.  If it is thought of as applying to a single system,
or a single universe, then at least from the perspective adopted in this paper,
it does not make much (quantum mechanical) sense.  It is only fair to add that
Dowker and Kent themselves show very little enthusiasm for their ``many
histories'' interpretation .

\section*{Acknowledgments}

	Correspondence, and in some cases conversations, with B.~d'Espagnat,
J.~Hartle, A.~Kent, and R.~Omn\`es have been very useful both in the
formulation of the ideas presented here, and in revising their presentation.
Financial support for this research was provided by the National Science
Foundation through grant PHY-9602084.

		\section*{App. A. Consistent Histories and Contrary Inferences}

	In recent work \cite{kt97,kt97b}, Kent has stated that the consistent
history approach cannot be taken seriously as a fundamental theory because it
allows for what he calls ``contrary inferences''.  We shall show that the
problem is not with consistent histories quantum theory as presented in CHQR
and the present paper, but rather with Kent's definition of ``contrary'', which
fails to take proper account of quantum incompatibility when it arises from
violations of consistency conditions.  For a condensed version of these
remarks, see \cite{gh97}.

	The Aharonov-Vaidman \cite{av91} example discussed in CHQR Sec.~VI~D
can be used to illustrate the central point of Kent's argument.  A particle can
be in one of three states $|A{\rangle}$, $|B{\rangle}$, or $|C{\rangle}$, and
the dynamics is trivial: $|A{\rangle}
\mapsto 
|A{\rangle}$, etc.  Define
\begin{equation} 
|\Phi{\rangle} = (|A{\rangle} + |B{\rangle} +
|C{\rangle})/\sqrt 3,\quad 
|\Psi{\rangle} = (|A{\rangle} + |B{\rangle} -
|C{\rangle})/\sqrt 3, 	 	
\label{eA.1} 
\end{equation}
and  (consistent with previous notation)  let a letter outside a ket denote
the corresponding projector, and a tilde its complement, thus:
\begin{equation} 
 A=| A{\rangle}{\langle} A|,\quad \tilde A = I-A = B + C,
\label{eA.2} 
\end{equation}
etc. Define a consistent family $\A$ of histories starting with $\Phi$ at time
$t_0$, followed by $A$ or $\At$ at a later time $t_1$, and $\Psi$ at a still
later time $t_2$.  It is
straightforward to show, using this framework, that 
\begin{equation} 
 \Pr(A_1\vb \Phi_0\land\Psi_2) = 1,
\label{eA.3} 
\end{equation}
where the subscripts indicate the time associated with the corresponding event.
That is, we can be sure that the particle was in state $A$ at $t_1$, given the
initial state $\Phi$ at $t_0$ and the final state $\Psi$ at $t_2$.  An
alternative framework $\B$, incompatible with $\A$, uses the same events at
$t_0$ and $t_2$, and $B$ and $\Bt$ at $t_1$.  Using $\B$, one finds
\begin{equation} \Pr(B_1\vb \Phi_0\land\Psi_2) = 1.
\label{eA.4} 
\end{equation}

	Kent defines two projectors $A$ and $B$ to be ``contrary''  provided
\begin{equation}
  AB=BA=0,\quad A\neq \Bt,
\label{eA.5}
\end{equation}
and employs the term ``contradictory'' when $A=\Bt$.  Since $A$ and $B$ in the
Aharonov-Vaidman example satisfy (\ref{eA.5}), Kent would conclude that
(\ref{eA.3}) and (\ref{eA.4}) are two probability-one inferences based on the
same data, $\Phi_0\land\Psi_2$, to two contrary events, and he finds this
feature of the consistent history approach to be problematical.

	To analyze this argument, we first note that the term {\it contrary}
has a well-defined usage in classical logic \cite{cn93}, where it indicates the
relationship between two propositions which cannot both be true, but might both
be false.  ``The queen is in London'' and ``the queen is in Cambridge'' are
contrary propositions in this sense: they are mutually exclusive, so they
cannot both be true, but they could both be false (if the queen is in some
other city).  Similarly, {\it contradictory} is reserved for the relationship
of two propositions which cannot both be true, and also cannot both be false.
For example, ``the queen is in London'' is false if and only if ``the queen is
not in London'' is true, and vice versa.

	Translating terms from classical logic into appropriate quantum
counterparts is not a trivial exercise; see CHQR and Sec.~\ref{incomp} of the
present paper.  As long as $A$ and $B$ belong to the same consistent family,
(\ref{eA.5}) is a reasonable quantum counterpart for the classical term
``contrary'', and agrees with the rules worked out in CHQR.  The problem with
Kent's argument is that he wishes to apply the same definition in a case in
which $A$ and $B$ do not belong to the same consistent family. In the
Aharonov-Vaidman example, $B$ is {\it incompatible}, in the quantum mechanical
sense (Sec.~\ref{incomp}) with the family $\A$ used to obtain (\ref{eA.3}),
while $A$ is incompatible with the family $\B$ used to obtain (\ref{eA.4}).
When quantum incompatibility obtains, the consistent history approach, for
reasons indicated in Sec.~\ref{incomp}, disallows any logical comparison
whatsoever.  Thus as long as one is considering inferences based upon the data
$\Phi_0\land\Psi_2$, $A$ and $B$ are best thought of as ``incomparable'', and
speaking of them as ``contrary'' in a sense similar to that used in classical
logic is misleading.

	To put the matter in another way, in order to be able to say that $A$
and $B$ are contrary in the sense of classical logic, consistent history rules
require that they belong to the same framework.  However, any framework which
contains both $A$ and $B$ at $t_1$ cannot also contain both $\Phi$ at $t_0$ and
$\Psi$ at $t_2$.  Consequently, in a framework in which it would be correct to
say that $A$ and $B$ are contrary, neither of the inferences (\ref{eA.3}) and
(\ref{eA.4}) is possible.  So, whichever way one looks at the matter, there are
no ``contrary inferences''.  The key point is that consistent history
reasoning, as clearly stated by Omn\`es \cite{n7} and reiterated in CHQR, must
employ a single framework, and Kent's argument violates this rule through
defining \cite{n8} ``contrary'' in a manner which allows it to hold as a
relationship between propositions which are not members of a single framework.

	The same point can be made in a slightly different way.  In
Sec.~\ref{example} we exhibited two probability one inferences, in
(\ref{e2.18}) and (\ref{e2.19}), based upon identical conditions but carried
out in two incompatible families, $\F_3$ and $\F_2$.  The nonsensical result
(\ref{e2.20}) of combining these two inferences was blocked by the consistent
history rule that results deduced in incompatible families cannot be combined.
It is precisely this same rule which applies in the case of (\ref{eA.3}) and
(\ref{eA.4}). The only difference is that while a quantum comparison of $c$ and
$s$ from (\ref{e2.18}) and (\ref{e2.19}) is obviously nonsensical, because the
projectors do not commute, that of $A$ and $B$ as inferred in (\ref{eA.3}) and
(\ref{eA.4}) is not as transparently incorrect, since the projectors commute,
even though one has an equally serious violation of precisely the same consistent
history rule.  Note that quantum incompatibility can arise in the consistent
histories approach both because certain projectors do not commute {\it and}
because consistency rules are violated.  The incompatibility of $A$ with $\B$,
and of $B$ with $\A$, is of the latter type, which is why it is a bit less
evident than the incompatibility of $c$ and $s$ in the example of
Sec.~\ref{example}. 

	In summary, to say that $A$ and $B$ are ``contrary'' in a logical sense
when they belong to incompatible frameworks is a violation of one of the basic
principles of consistent history reasoning, at least as the subject has been
developed up to now.  To be sure, there may exist alternative approaches to
consistent histories based upon a different sent of logical rules,
and one might view Kent's argument as, in effect, proposing such an
alternative.  In that case there would be no reason to disagree with his
conclusion, which would be that this alternative proposal constitutes a
formalism which cannot be taken seriously as a fundamental theory of nature,
for precisely the reasons which he points out.

	Finally, a comment on the question which Kent raises in the latter part
of \cite{kt97}: why should the formalism of consistent histories rule out
inferences to two ``contradictory'' propositions while allowing inferences to
two ``contrary'' propositions?  The brief response is that, according to the
formalism for consistent histories developed in CHQR, neither ``contrary'' nor
``contradictory'' can be defined as logical relationships unless both of the
properties (or histories) being compared are found in the same consistent
family.  Thus this formalism never allows an inference to either
``contradictory'' or ``contrary'' pairs of propositions, and so the question
raised by Kent does not arise.  Of course, alternative formulations of
consistent histories which construct logical definitions in a way which allows
contrary inferences to occur are subject to the conceptual problem which
Kent has pointed out, and have to deal with it in some way.  For one such
alternative, see \cite{kt97b}

%\newpage

		\section*{App. B. Conditioning Upon Quasiclassicality}

	Kent has claimed that  even if one assumes that the world will be
quasiclassical tomorrow, the consistent history approach does not always yield
probabilistic predictions which agree with those provided by Copenhagen (i.e.,
standard textbook) quantum theory.  The following example illustrates his
argument, and shows why it cannot be considered a serious  objection to 
consistent histories as described  in CHQR and the present paper.

	Suppose the beamsplitter in the example in Sec.~\ref{example} (Fig~1)
is replaced by one which produces beams in three exit channels $c$, $d$, and
$e$, corresponding to a unitary time development
\begin{equation}
    | a\rg\mt (|c\rg + |d\rg +|e\rg)/\sqrt{3}
\label{eB.1}
\end{equation}
in place of (\ref{e2.1}).  A third detector $E$ is added for the $e$ channel,
so that 
\begin{equation}
  |e\rg|E\rg\mt |E^*\rg
\label{eB.2}
\end{equation}
in addition to (\ref{e2.2}).  We now consider two consistent families involving
two times, $t_0$ and $t_2$:
\begin{eqnarray}
  \D_1: &&\quad \Psi_0\od\{ C^*DE,\,CD^*E,\, CDE^*\},
\label{eB.3}\\
  \D_2: &&\quad \Psi_0\od\{ SE,\, CDE^*\},
\label{eB.4}
\end{eqnarray}
where the initial state (the counterpart of (\ref{e2.4})) is 
\begin{equation}
   |\Psi_0\rg = |aCDE\rg,
\label{eB.5}
\end{equation}
$|S\rg$ is the MQS combination of $|C^*D\rg$ and $|D^*C\rg$ defined in
(\ref{e2.5}), and the curly brackets in (\ref{eB.3})  and (\ref{eB.4}) enclose
alternative possibilities at $t_2$.  As usual, various histories of zero weight
have been omitted. 

	A straightforward analysis using $\D_1$ yields
\begin{equation}
   \Pr(C^*DE\vb\Psi_0)=\Pr(CD^*E\vb\Psi_0)=\Pr(CDE^*\vb\Psi_0) =1/3,
\label{eB.6}
\end{equation}
that is, the probability is 1/3 that each of the detectors will have detected
the photon, whereas from $\D_2$ one concludes that
\begin{equation}
  \Pr(SE\vb\Psi_0) = 2/3,\quad \Pr(CDE^*\vb\Psi_0) =1/3.
\label{eB.7}
\end{equation}
The fact that $\Pr(CDE^*\vb\Psi_0)$ is the same in both cases reflects a
general property of the consistent histories approach, as noted in
Sec.~\ref{example}.

	Kent would accept (\ref{eB.6}) and (\ref{eB.7}) as correct, but would
then argue that {\it if one conditions upon quasiclassicality}, the probability
that $E$ detects the photon is different in families $\D_1$ and $\D_2$; to be
specific:
\begin{eqnarray}
  \D_1: && \Pr(CDE^*\vb\Psi_0\land\mbox{quasiclassical}) =1/3,
\label{eB.8}\\
  \D_2: && \Pr(CDE^*\vb\Psi_0\land\mbox{quasiclassical}) =1.
\label{eB.9}
\end{eqnarray}
The argument is that all three
histories in $\D_1$ are quasiclassical (no MQS states), and thus each has a
probability 1/3, whether or not one conditions upon quasiclassicality.  On the
other hand, in $\D_2$ the only quasiclassical history is $\Psi_0\od CDE^*$,
since the other involves the far-from-classical superposition $S$, and thus
conditioning upon quasiclassicality yields (\ref{eB.9}) in place
of (\ref{eB.7}).

	The problem with this argument, if one adopts the point of view of
CHQR, is that conditional probabilities are only defined when their arguments
are projectors belonging to an appropriate Boolean algebra satisfying
consistency conditions.  But, as noted in Sec.~\ref{predict}, there is no
subspace of the Hilbert space, and hence no projector, corresponding to
``quasiclassical'', understood in the present context as ``non-MQS''.  Thus the
conditional probabilities (\ref{eB.8}) and (\ref{eB.9}) are undefined, and no
argument based upon them can be valid.

	Is there some way to define ``quasiclassical'' as a condition entering
into a probability without referring to a subspace of the Hilbert space?  If,
as suggested in Sec.~\ref{predict}, the term ``quasiclassical'' belongs to the
metalanguage rather than the language of quantum descriptions, such a
definition would change the meaning of the resulting probabilities in an
important way.  Thus imagine that $\D_1$ and $\D_2$ were two coarse grainings of
a classical phase space using cells of different sizes.  Then by conditioning
upon the initial condition and the fact that the history only involves, say,
large cells, one could produce a difference analogous to that between
(\ref{eB.8}) and (\ref{eB.9}).  But the result would obviously have no direct
physical significance, although it might be useful in addressing the question
of which coarse graining is most useful for a numerical simulation.

	Of course, there may be some alternative to CHQR in which
``quasiclassical'' refers directly to a quantum property. In that case, Kent's argument would indicate that this alternative (in contrast
to CHQR) is not likely to reproduce standard quantum physics.

\vspace{2 cm}

	FIGURES ON NEXT PAGE
\newpage

\epsfxsize=15truecm
\epsfbox{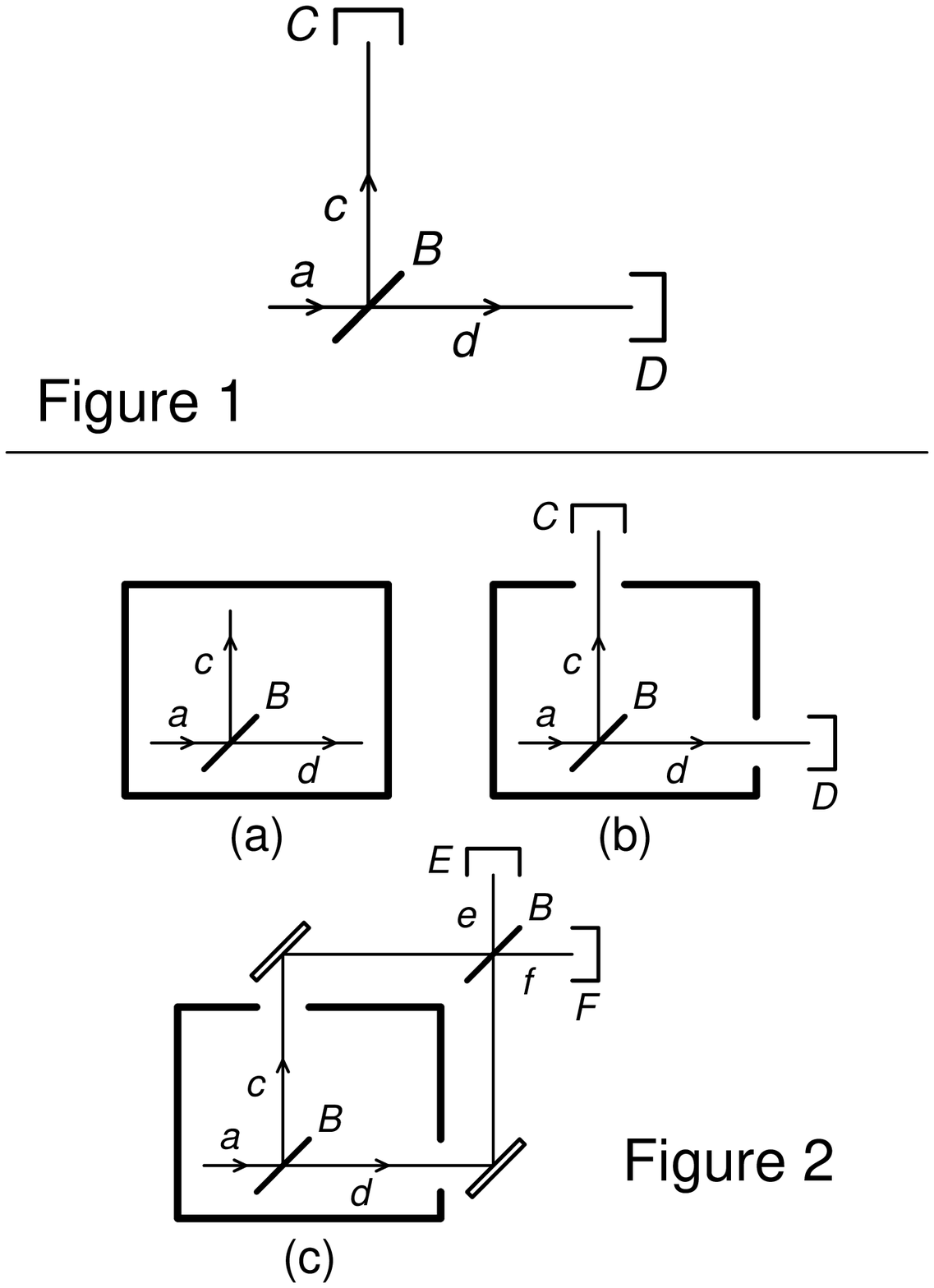}

Fig. 1. Beamsplitter $B$ and detectors $C$ and $D$.

Fig. 2. Histories for the closed system in (a) can be confirmed using either
the configuration shown in (b) or the one in (c).


\begin{thebibliography}{11}
\bibitem{n3}
	One sometimes encounters discussions of how one can ``measure'' whether
$S_x=1/2$ or $S_z=1/2$.  From the consistent history perspective, this is
somewhat careless terminology; what the writers have in mind is, typically, a
situation in which a macroscopic apparatus can be set to produce a spin half
particle either in the state $S_x=1/2$ or in the state $S_z=1/2$.  Because the
settings represent macroscopically distinct situations, the two apparatus
states (viewed as quantum states) are orthogonal, and the measurement on the
spin half particle is then used to try and infer something about the apparatus
states.
\bibitem{n3b}
	Even in a classical world it would be impossible for a physical system
to contain a complete description of itself.  Consequently, it is always
simplest to imagine that a quantum description is being constructed outside the
system which it purports to describe.  Sometimes the question is asked as to
how we can describe, in a scientific sense, the universe in which we live.
This is a non-trivial philosophical issue, but it does not seem implausible to
assume that some sort of analogy will work: we can imagine simple model systems
inside a closed box, check that they have some features which we find in our
everyday experience, and then extrapolate by imagining ourselves in a large
box, etc.
\bibitem{n8}
	One way to avoid confusion would be to introduce a new term, say
``perp'', to denote the relationship (\ref{eA.5}) between two projectors.
If  $A$ and $B$ are ``perp'' and belong to the same consistent family,
this means that the corresponding propositions are ``contrary'' in the
logical sense.  But if  $A$ and $B$ are not in the same consistent family,
``perp'' says nothing about a logical relationship.
\end{thebibliography}
\end{document}